\documentclass[
aps,pra,
amsmath,
amssymb,
amsfonts,
10pt,
showpacs,
twocolumn,
superscriptaddress
]{revtex4-1}

\usepackage{braket}

\usepackage{qcircuit}

\usepackage{graphicx} 
\usepackage[caption=false]{subfig}

\usepackage{hyperref}

\usepackage{xcolor}

\newcommand{\romnum}[1]{\textcolor{gray}{#1}}

\usepackage{xspace}
\newcommand{\equref}{Eq.\xspace}
\newcommand{\figref}{Fig.\xspace}
\newcommand{\refref}{Ref.\xspace}

\newcommand{\appref}{Appendix\xspace}
\newcommand{\etal}{\textit{et al.}\xspace}

\begin{document}


\title{Variational quantum algorithms for discovering Hamiltonian spectra}


\author{Suguru Endo \& Tyson Jones}
\affiliation{Department of Materials, University of Oxford, Parks Road, Oxford OX1 3PH, United Kingdom}



\author{Sam McArdle}
\affiliation{Department of Materials, University of Oxford, Parks Road, Oxford OX1 3PH, United Kingdom}
\author{Xiao Yuan}
\affiliation{Department of Materials, University of Oxford, Parks Road, Oxford OX1 3PH, United Kingdom}
\author{Simon Benjamin}
\affiliation{Department of Materials, University of Oxford, Parks Road, Oxford OX1 3PH, United Kingdom}


\begin{abstract}
Calculating the energy spectrum of a quantum system is an important task, for example to analyse reaction rates in drug discovery and catalysis. There has been significant progress in developing algorithms to calculate the ground state energy of molecules on near-term quantum computers. However, calculating excited state energies has attracted comparatively less attention, and it is currently unclear what the optimal method is. We introduce a low depth, variational quantum algorithm to sequentially calculate the excited states of general Hamiltonians. Incorporating a recently proposed technique~\cite{Excited}, we employ the low depth swap test to energetically penalise the ground state, and transform excited states into ground states of modified Hamiltonians. We use variational imaginary time evolution as a subroutine, which deterministically propagates towards the target eigenstate. We discuss how symmetry measurements can mitigate errors in the swap test step. We numerically test our algorithm on Hamiltonians which encode 3SAT optimisation problems of up to 18 qubits, and the electronic structure of the Lithium Hydride molecule.
As our algorithm uses only low depth circuits and variational algorithms, it is suitable for use on near-term quantum hardware. 
\end{abstract}

\maketitle


\section{Introduction}

Many physical properties of a quantum system are determined primarily by its energy spectrum. Diagonalisation of the Hamiltonian allows one to calculate various expectation values and correlation functions \cite{classical_diag_hamils}.  
For example, the energy spectra of molecules inform their dynamics and therefore an understanding of such spectra is vital for molecular design~\cite{diag_in_chemistry_example}.
But diagonalising the Hamiltonians of quantum systems on a classical machine is an often intractable task. The exponentially growing cost of storing and operating upon the quantum system makes diagonalising large systems prohibitively expensive. 
This precludes, for example, the study of complicated compounds~\cite{classical_chem_diag_expensive}.

It is widely believed that quantum computers will make these classically intractable molecular simulations possible~\cite{Revolution}. This was formalised by Aspuru-Guzik~\etal, who suggested using the adiabatic state preparation and phase estimation algorithms to find the ground state energy of molecules~\cite{aspuru2005simulated}. Such a method necessitated deep quantum circuits and therefore long coherence times. The recently proposed variational quantum eigensolver (VQE) circumvents these requirements, exchanging them for an increased number of circuit repetitions~\cite{peruzzo2014variational,VQETheoryNJP}. To date, there have been several proof of principle experiments which have applied the VQE to find the ground state energy of small molecules~\cite{peruzzo2014variational,kandala2017hardware,TrappedIon,PRXH2,wang2015quantum}. Other variational algorithms have been introduced which can simulate the real~\cite{Li2017} or imaginary~\cite{mcardle2018variational} time dynamics of quantum systems. In particular it was shown that imaginary time evolution can be used as an alternative to the VQE to find the ground state of molecular Hamiltonians. 

While the ground state problem has received significant attention, the problem of finding the excited states of molecular systems has experienced comparatively less development.  
This is despite its particular importance in analysing chemical reactions, which is a vital ingredient in the quest to discover new drugs and industrial catalysts~\cite{Reiher201619152}.

There have thus far been a handful of proposals for calculating excited states, all based on the VQE, such as the quantum subspace expansion method~\cite{subspace1,subspace2} and the von-Neumann entropy method~\cite{Santagatieaap9646}. These methods require either many measurements, or deep quantum circuits, for instance to perform quantum phase estimation.

In this work, we propose a variational algorithm which uses imaginary time evolution to sequentially calculate the energy levels of a Hamiltonian. The algorithm makes use of the shallow swap test~\cite{shallowswap1,shallowswap2} to evaluate the overlap of two input wavefunctions~\cite{Excited}. We first use imaginary time evolution to target the ground state of the Hamiltonian. 
Using the shallow swap test, we can energetically penalise the ground state wavefunction, then discover the other eigenstates through repeated evolution and penalisation. This method makes use of only shallow circuits, at the cost of additional measurements.

A recent work by Higgott~\etal~\cite{Excited} introduces the use of the swap test with the VQE to discover the energy eigenstates of the diatomic Hydrogen molecule. 
Here, we contrast the performance of methods based on direct descent with our imaginary time approach, finding that the former is prone to becoming stuck in non-physical local minima of the parameter space~\cite{PhysRevA.92.042303,mcardle2018variational}. This may render them unsuitable for probing the full spectra of bigger systems. We numerically demonstrate this for a 6-qubit molecular Hamiltonian.

Conversely, we find that when evolution is restricted to a submanifold of the full Hilbert space, variational imaginary time evolution tends to converge to energy eigenstates of the Hamiltonian, regardless of the initial state. This is a crucial mechanism exploited by our algorithm to reliably penalise and discover the physical energy spectrum. We test our method on Hamiltonians which encode the boolean satisfiability problem (3SAT), and to find the electronic spectrum of the Lithium Hydride (LiH) molecule.


\section{Imaginary time evolution}

Our algorithm makes use of variational imaginary time evolution, to be performed by a hybrid quantum-classical machine. We briefly outline the procedure below. See
\refref~\cite{mcardle2018variational} for a more detailed discussion.

For a time independent Hamiltonian, $H$, the normalised imaginary time evolution is given by 
\begin{equation}
\begin{aligned}
\ket{\psi(\tau)}=\frac{e^{-H \tau} \ket{\psi(0)}}{\sqrt{\bra{\psi(0)} e^{-2H \tau} \ket{\psi(0)}}},
\label{imaginary}
\end{aligned}  
\end{equation}
which is the solution of the imaginary time Schr\"odinger equation 
\begin{equation}
\begin{aligned}
\frac{d \ket{\psi(\tau)}}{d\tau}= (H-\braket{H(\tau)}) \ket{\psi(\tau)},
\end{aligned}
\end{equation}
where $\braket{H(\tau)}=\bra{\psi(\tau)} H\ket{\psi(\tau)}$. 
Forgoing normalisation, a general state $\ket{\psi} = \sum_j c_j \ket{e_j}$ evolves in imaginary time like
\begin{align}
\ket{\psi(\tau)} \sim \sum \limits_j c_j e^{-E_j \tau} \ket{e_j},
\end{align}
where the probability of energy eigenstates $\ket{e_j}$ decay exponentially with their energies $E_j$.
Provided that $\ket{\psi(\tau)}$ has a non-zero overlap with the ground state $\ket{g}$, it can be verified that $\lim_{\tau \rightarrow \infty} \ket{\psi(\tau)} =\ket{g}$.
While non-unitary imaginary time evolution cannot be directly implemented on a quantum computer, it can be simulated using a hybrid quantum-classical algorithm.
The state $\ket{\psi(\tau)}$ is approximated by a parametrized trial state $\ket{\varphi (\theta_1 (\tau), ... , \theta_M(\tau) )}:= \ket{\varphi (\vec{\theta}(\tau) )}$, and its evolution determined by the evolution of $\vec{\theta} (\tau)$. 
The trial state is produced by an ansatz quantum circuit $ \ket{\varphi (\vec{\theta}(\tau) )} = U(\theta_M) U(\theta_{M-1})...U(\theta_{1}) \ket{\bar{0}} $, where $U(\theta_k (\tau))$  is in practice a single or two qubit gate.

The evolution of the parameters $\vec{\theta} (\tau)$ under imaginary time evolution is given by
\begin{equation}
\begin{aligned}
\sum_j \mathcal{M}_{ij} \dot {\theta}_j &= \mathcal{V}_i, 
\label{update}
\end{aligned} 
\end{equation}
where 
\begin{equation}
\begin{aligned}
\mathcal{M}_{ij} &= \Re \bigg{(}\frac{\partial \bra{\varphi(\vec{\theta}(\tau))}}{\partial \theta_i} \frac{\partial \ket{\varphi(\vec{\theta}(\tau))}}{\partial \theta_j}\bigg{)}, \\
\mathcal{V}_i&= \Re \bigg{(} \bra{\varphi(\vec{\theta}(\tau))} H \frac{\partial \ket{\varphi(\vec{\theta}(\tau))}}{\partial \theta_i}\bigg{)}. 
\end{aligned}
\label{eq:m_and_v_update_equs}
\end{equation}
These elements are obtained by the quantum processor using the shallow quantum circuit shown in \appref~\ref{appendix:eval_M_V}. 
The classical processor can then update the parameters using the Euler update rule
\begin{equation}
\begin{aligned}
\vec{\theta}(\tau +\delta \tau)= \vec{\theta} (\tau) + \delta \tau \mathcal{M}^{-1} \mathcal{V}.
\end{aligned}
\label{eq:update_theta_eq}
\end{equation}
If the ansatz is sufficiently powerful, repeatedly constructing and solving this linear equation will evolve the system to a state close to the ground, which we denote as $\ket{\tilde{g}}$.
We monitor convergence by the change in the parameters, and halt when $\| \Delta \vec{\theta}(\tau) \| = \|\delta \tau \mathcal{M}^{-1} \mathcal{V} \| \approx 0$.
The expected energy of a converged state is easily evaluated using a polynomial number of simple Pauli operators \cite{peruzzo2014variational}.

With a less powerful ansatz, imaginary time evolution may fail to reach the ground state, but tends to converge to a higher excited eigenstate of the Hamiltonian. We do not presently provide a proof of this, though this behaviour is seen consistently in our numerical simulations. 



\section{Evaluation of the energy spectrum of the Hamiltonian using imaginary time evolution}

We now describe how to evaluate the excited states of the Hamiltonian.
Having found an approximate ground state $\ket{\tilde{g}}$, we can construct a modified Hamiltonian 
\begin{align}
H' = H + \alpha \ket{\tilde{g}}\bra{\tilde{g}},
\end{align}
which, for sufficiently large $\alpha \in \mathbb{R}$, no longer has ground state $\ket{g}$. Instead, the first excited state $\ket{e_1}$ of $H$ becomes the ground state of $H'$, and $\ket{\tilde{g}}$ is now an excited state of $H'$ with energy increased by $\alpha$, which will decay exponentially faster in imaginary time. The rest of the spectrum, orthogonal to $\ket{\tilde{g}}$, is unaffected.
A system evolving under $H'$ in imaginary time will then approach $\ket{e_1}$ instead. This state can in turn be excited, and the system evolved under Hamiltonian
\begin{align}
H'' = H + \alpha \ket{\tilde{g}}\bra{\tilde{g}} + \alpha  \ket{\tilde{e_1}}\bra{\tilde{e_1}}
\end{align}
to reach the next excited state of the original Hamiltonian. 
We can repeat this process by preparing the effective Hamiltonian $H+\alpha\ket{\tilde{g}} \bra{\tilde{g}}+\sum_{j=1}^N \alpha \ket{\tilde{e}_j}\bra{\tilde{e}_j}$ to obtain the $(N+1)$\textsuperscript{th} excited state $\ket{\tilde{e}_{N+1}}$.
In principle, we can obtain the complete energy spectrum, including a count of the degeneracies, so long as $\alpha$ is kept greater than the gap between ground and the highest state sought.
Note the order of the discovered and subsequently excited eigenstates is unimportant.

In practice we do not directly modify the Hamiltonian, as doing so would require full tomography of the state vector, which is exponentially costly. Instead, we modify the imaginary time evolution equations to describe the evolution under the modified Hamiltonian, $H'$. We replace $\mathcal{V}$ by
\begin{equation}
\begin{aligned}
\mathcal{V}_i= \Re\bigg(\frac{\partial \bra{\varphi(\vec{\theta}(\tau))}}{\partial \theta_i} &H \ket {\varphi(\vec{\theta}(\tau))} + \\
			&\alpha \frac{\partial \bra{\varphi(\vec{\theta}(\tau))}}{\partial \theta_i}  \ket{\tilde{g}} \braket{\tilde{g} |\varphi(\vec{\theta}(\tau))} \bigg),
\end{aligned} 
\label{eq:v_update_under_excitation}
\end{equation}
and so on to excite all discovered eigenstates by $\alpha$.

These additional terms to $\mathcal{V}_i$ can be evaluated using the low depth swap test circuit \cite{shallowswap2,shallowswap1}, outlined in \appref~\ref{appendix:overlap_calc}. We use the swap test to evaluate terms $|\braket{\varphi(\theta_i+\delta \theta_i)|\tilde{g}}|^2$ and $|\braket{\varphi|\tilde{g}}|^2$, and then approximate
\begin{gather}
\Re\bigg( \frac{\partial \bra{\varphi}}{\partial \theta_i}  
 \ket{\tilde{g}} \braket{\tilde{g} |\varphi} \bigg)
 =
 \frac{1}{2} \frac{\partial}{\partial \theta_i} | \braket{ \varphi | \tilde{g} } |^2 \tag*{}
 \\
 \simeq
 \frac{1}{2}
 \frac{|\braket{\varphi(\theta_i+\delta \theta_i)|\tilde{g}}|^2-|\braket{\varphi|\tilde{g}}|^2}{\delta \theta_i}
\end{gather}
for some sufficiently small $\delta \theta_i$.

There is no requirement that each discovered eigenstate is excited by the same amount in the modified Hamiltonian. We may vary $\alpha$ for each excited state. In that scenario, one can add
\begin{align}
\sum \limits_j^N \alpha_j \,\Re\bigg( \frac{\partial \bra{\varphi(\vec{\theta}(\tau))}}{\partial \theta_i}  
 \ket{\tilde{e_j}} \braket{\tilde{e_j} |\varphi(\vec{\theta}(\tau))} \bigg)
 \label{eq:updating_V_with_different_excitations}
\end{align}
to $\mathcal{V}_i$ in Eq.~\eqref{eq:m_and_v_update_equs} to emulate a Hamiltonian with energy eigenvalues $\{ E_1+\alpha_1,\;\dots, \; E_N + \alpha_N \}$.

\section{Error mitigation}

\begin{figure}[htbp!]
\includegraphics[width=8cm]{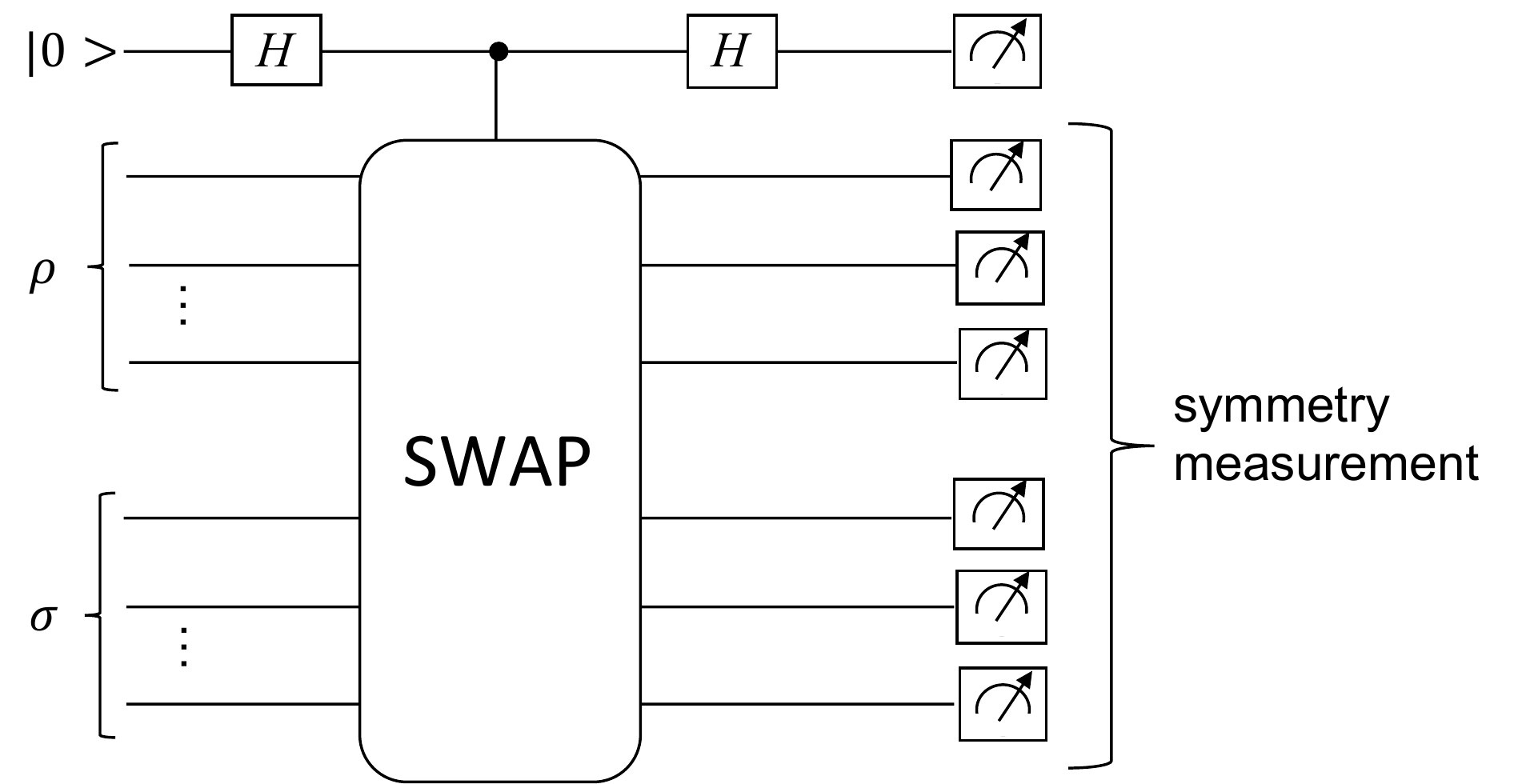}
\caption{Using symmetry measurements to detect and mitigate errors in the conventional swap test.}
\label{stest2}
\end{figure}

We raise the possibility of applying error mitigation to our algorithm, through a simple error detection routine. Instead of using the low depth swap test described above, we can also use the conventional swap test (Fig. \ref{stest2}~\cite{nielsen2002quantum}). The depth of this circuit grows linearly with the number of qubits used. The conventional swap test calculates the overlap between the two states by measuring an ancilla. However, no measurements are performed on the register qubits, and so any information we gain from them is, in a sense, free. We consider the input states as $\ket{g}, \ket{e}$. After the conventional swap test circuit, the register is left in the state
\begin{align}
\ket{\phi_\mathrm{r}^{\pm}}=\frac{1}{\sqrt{2}}(\ket{g}\ket{e}\pm\ket{e}\ket{g})
\label{symmetry}
\end{align}
where the sign is determined by the measurement result of the ancilla qubit. The state $\ket{\phi_\mathrm{r}^{\pm}}$ will be invariant under a symmetry $S$, if both $\ket{g}$ and $\ket{e}$ are also invariant under $S$. If we make a measurement of this symmetry on the register, we will be able to detect errors which break this symmetry. We can then discard those results for which we detect an error.

In the case of molecular Hamiltonians, we are often interested in ground and excited states which conserve the number of electrons in the molecule. If we use an ansatz which conserves the number of electrons (such as the unitary coupled cluster ansatz~\cite{romero2017strategies}) then a measurement of the electron number operator, $\hat{N}_e$, on the output state of the swap test should give the total number of electrons. If it does not, then an error has occurred, and we can discard the measurement. This method can thus mitigate the effect of single qubit bit flip errors, and certain combinations of two qubit errors. This error mitigation method can also be applied to the method developed in Ref.~\cite{Excited}. 

Moreover, our algorithm is compatible with the other error mitigation techniques proposed in Refs.~\cite{endo2017practical,PhysRevLett.119.180509}.
We do not test these strategies in the present work.


\section{Numerical simulations}

\begin{figure}[b]
\centering
\includegraphics[width=1\columnwidth]{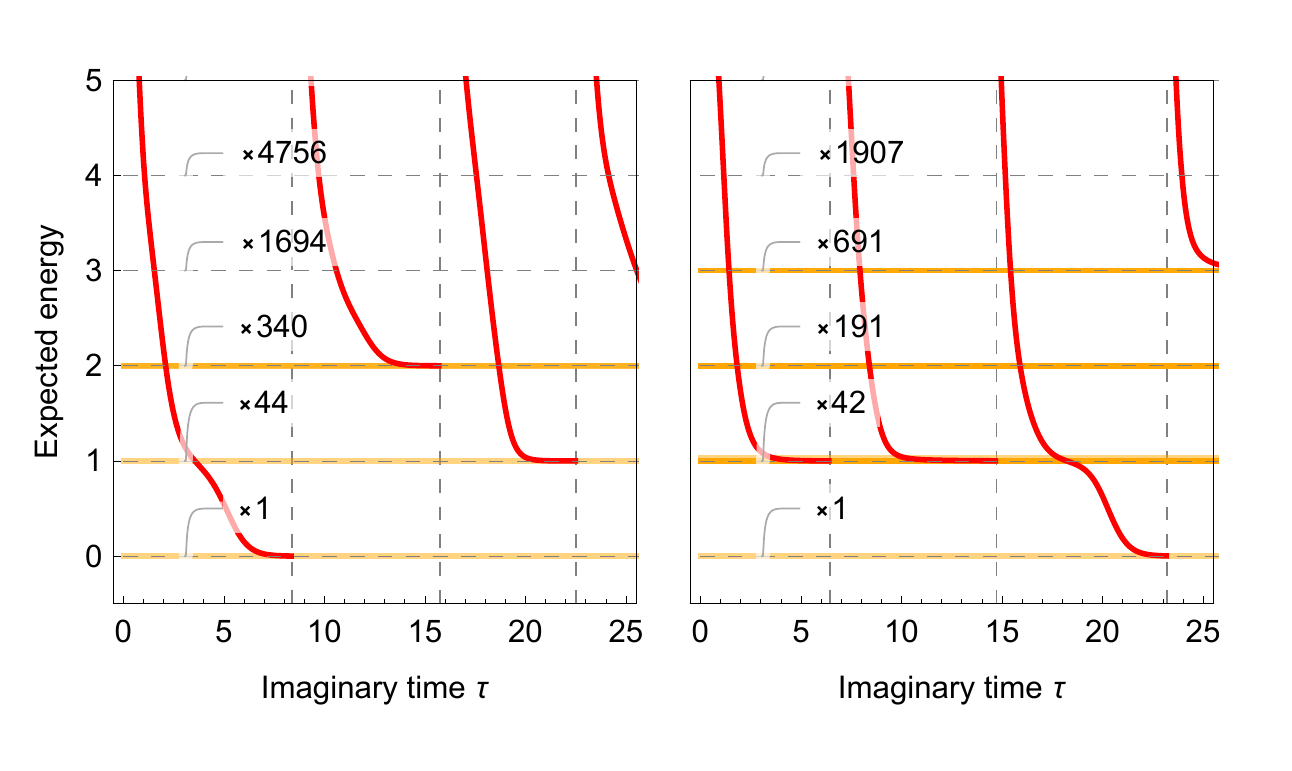}
\caption{The expected energy as variational imaginary time evolution discovers some low lying states of 18 boolean (left) and 16 boolean (right) 3SAT Hamiltonians, using the compact ansatz (with 126 and 112 parameters respectively). Vertical dashed lines indicate iterations when the Hamiltonian was excited and the parameters re-randomised. Horizontal dashed and coloured lines indicate the true eigenvalues and those found by our method respectively. Labels indicate the degeneracies of the states.}
\label{fig:3sat_example}
\end{figure}

\begin{figure*}[htb]
\centering
	
\subfloat[t][] {
	\centering
	\includegraphics[width=\columnwidth]{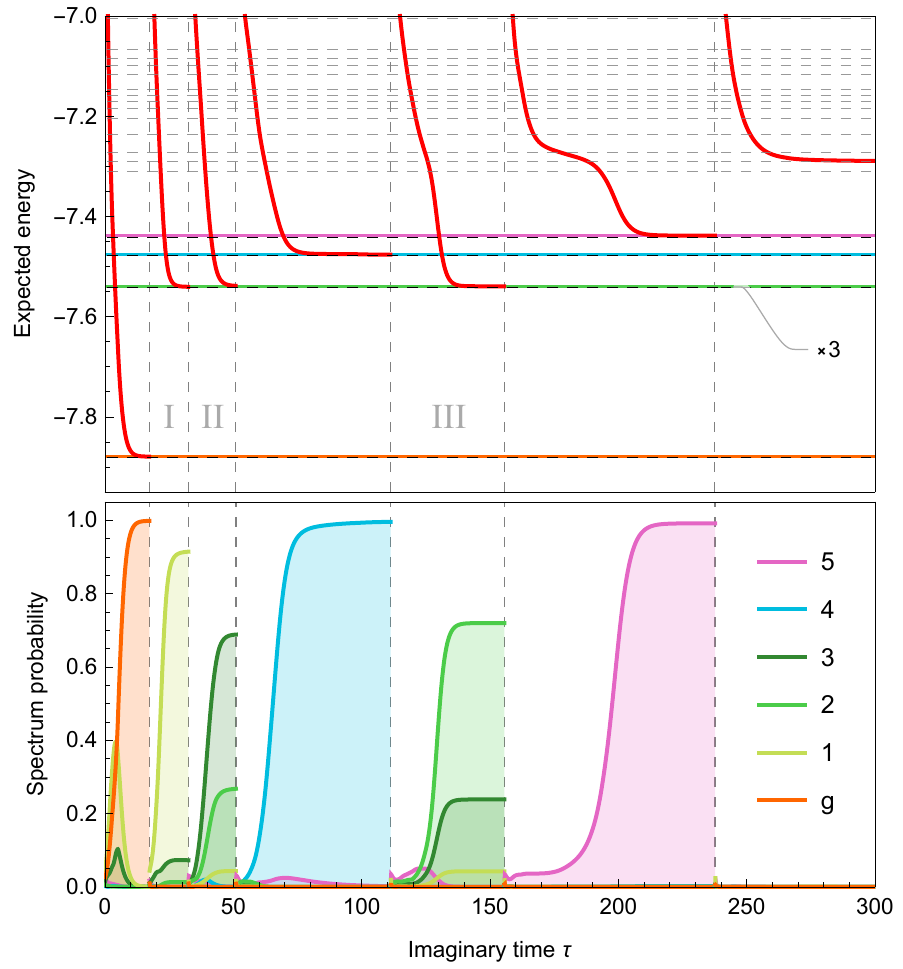}
	\label{fig:eigs_imagtime_example}
} %
\subfloat[t][
    	] {
	\centering
	\includegraphics[width=\columnwidth]{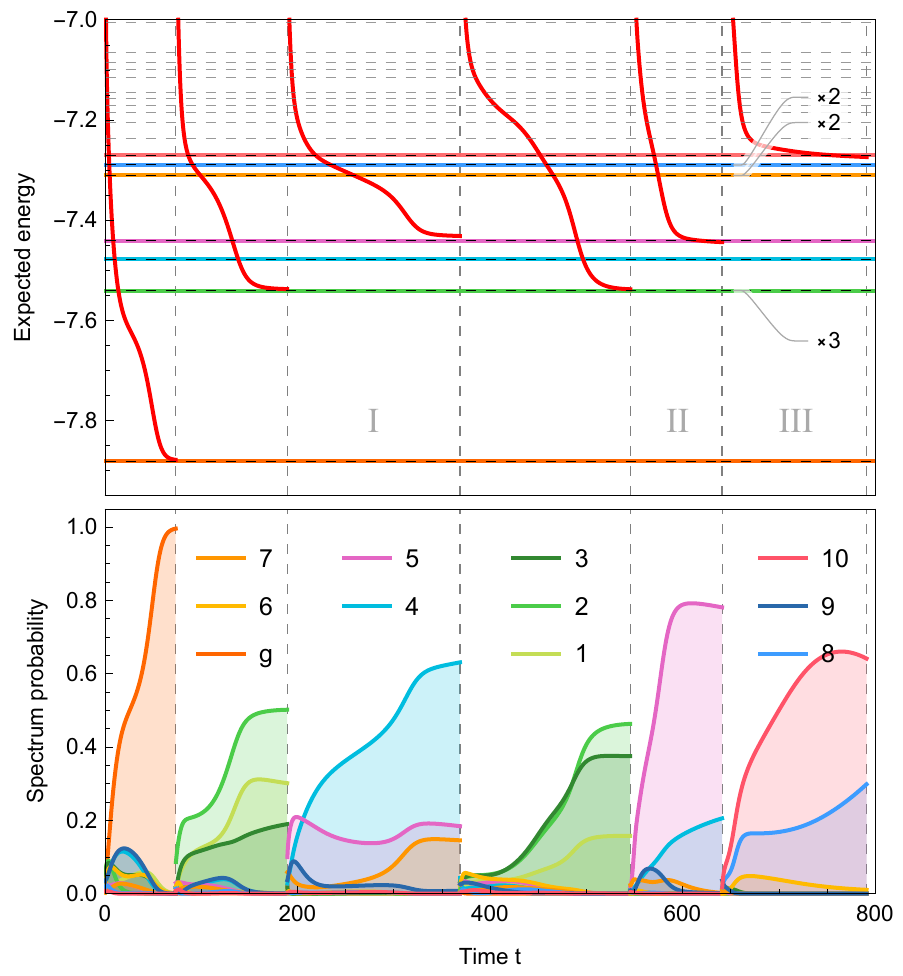}
	\label{fig:eigs_graddesc_example}
}
\caption{Variational simulations discovering then exciting several low lying energy eigenstates of the reduced 6-qubit LiH Hamiltonian, using the low depth ansatz with 56 parameters. The top plot shows the expected energy in red, as the states reached at the vertical dashed lines are excited in the Hamiltonian. Horizontal dashed and coloured lines indicate the true and discovered energy eigenvalues respectively.
		The bottom plot shows the population of the eigenstates as found by numerically diagonalising the Hamiltonian.
        The spectrum discovered in the long-term is included in \figref~\ref{fig:specfound_6qb}. (a) \textbf{Imaginary time}. Regions \romnum{I}, \romnum{II} and \romnum{III} converge to orthogonal superpositions of the three degenerate first-excited states, which are themselves eigenstates.
	(b) \textbf{Gradient descent}.
		The green (labelled 1, 2 and 3), orange (labelled 6 and 7) and blue (labelled 8 and 9) states are degenerate.
		Regions \romnum{I}, \romnum{II} and \romnum{III} show gradient descent converging to non-eigenstates.}
\end{figure*}

We numerically simulate our algorithm with several Hamiltonians and several ans\"atze.
Each time, our initial parameter values are random, and parameters are re-randomised when we excite states in the Hamiltonian. We employ Tikhonov regularisation when updating the parameters to ensure smoothness. 
We elaborate on these details and further describe our numerical methods in \appref~\ref{appendix:numerics_details}.

The choice of ansatz is very important in variational simulation, and in this work, we explore the use of two.
We try the recently proposed \textit{low depth} circuit ansatz (LDCA)~\cite{dallaire2018low} which is chemically motivated and was found to outperform the unitary coupled cluster ansatz for the molecule cyclobutadiene. We also employ the ansatz recently used in Ref.~\cite{mcardle2018variational} to find the ground-state with imaginary time, which we refer to as the \textit{compact ansatz}. Both ans\"atze scale linearly with the number of qubits, and can be considered hardware efficient~\cite{dallaire2018low,mcardle2018variational}.

We task our algorithm with finding the spectra of simple Hamiltonians which encode the 3SAT optimisation problem, and more complicated Hamiltonians which encode the electronic structure of LiH. 
We describe their structure below - see \appref~\ref{appendix:hamil_construction} for a detailed descrition of their construction.
The 3SAT Hamiltonians are diagonal in the classical basis;
\begin{align}
H = \sum \limits_j n_j \ket{j}\bra{j}
\end{align}
where $n_j$ is the number of 3SAT clauses violated by the $j^\text{th}$ classical state when treated as a boolean assignment.
This yields equally-spaced, highly degenerate spectra.
We consider 3SATs Hamiltonians of up to 18 qubits.
The LiH Hamiltonian can be simplified by employing various physical approximations - we do this to reduce the full 12 qubit Hamiltonian to 10 and 6 qubit representations.

Molecular Hamiltonians can be written as a linear combination of products of local Pauli operators,
\begin{equation}\label{VQE_Hamil}
	H = \sum_j^M h_j\prod_i \sigma_i^j,
\end{equation}
where $\sigma_i^j$ represents one of $I$ , $\sigma^x$ , $\sigma^y$  or $\sigma^z$, $i$ denotes which qubit the operator acts on, and $j$ denotes which term in the Hamiltonian we apply. For example
\begin{equation}\label{VQE_Hamil_terms}
	H = h_0I + h_1 X_0 Y_1 Z_5 + h_2 Z_0 Y_3 Y_5 + \cdots
\end{equation}

We compare the spectrum reported by our simulated method with the eigenvalues of these Hamiltonian as found by exact numerical diagonalisation.



In \figref~\ref{fig:3sat_example}, we present a simulation of our method exploring the simple spectrum of some 3SAT problems. The vertical axis is the expected energy $\braket{\varphi(\vec{\theta}(\tau)) | H | \varphi(\vec{\theta}(\tau))}$ of the ansatz state, which we note is not necessary to monitor experimentally. The expected energy monotonically decays under imaginary time evolution until the system converges into an eigenstate, which is subsequently excited. It is interesting to note that the ground state is not necessarily discovered first, as demonstrated by the 16 qubit (right) simulation in \figref~\ref{fig:3sat_example}, where ground is the \textit{third} state discovered.

\begin{figure}[t]
		\centering
		\includegraphics[width=\columnwidth]{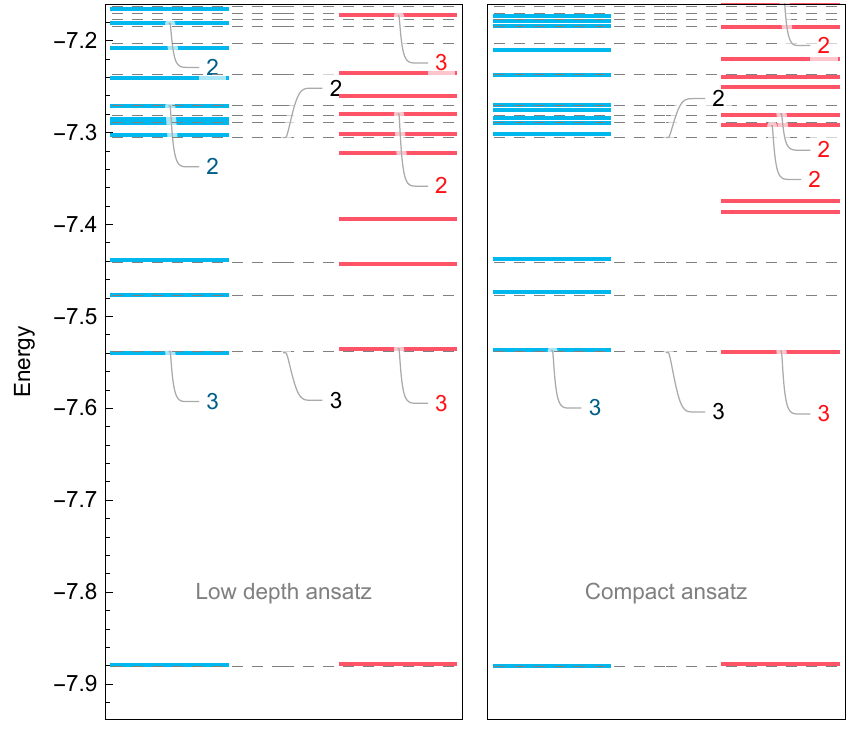} \\
		\vspace{5pt} \hspace{20pt} 
		\includegraphics[width=0.85\columnwidth]{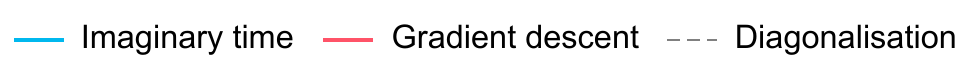}
		\caption{
			The $6$-qubit LiH spectra discovered by a single process of parameter evolution using imaginary time and gradient descent, compared to that found by direct numerical diagonalisation of the Hamiltonian. The low depth and compact ans\"atze use 56 and 42 parameters respectively. The low depth ansatz simulations extend those in Figs.~\ref{fig:eigs_imagtime_example} and \ref{fig:eigs_graddesc_example}. Energies closer than $5 \times 10^{-3}$ apart are combined and their degeneracy labelled.
	}
		\label{fig:specfound_6qb}
	\end{figure}

For the more complicated reduced $6$-qubit LiH Hamiltonian, we show that the variational imaginary time evolution successfully discovers eigenstates in 
\figref~\ref{fig:eigs_imagtime_example}.
In contrast, \figref~\ref{fig:eigs_graddesc_example} reveals gradient descent converging to non-eigenstates which when subsequently excited, modify the Hamiltonian in a non-trivial way. This leads to errors in the discovered spectrum, shown in \figref~\ref{fig:specfound_6qb}.

\begin{figure}[t]
	\includegraphics[width=\columnwidth]{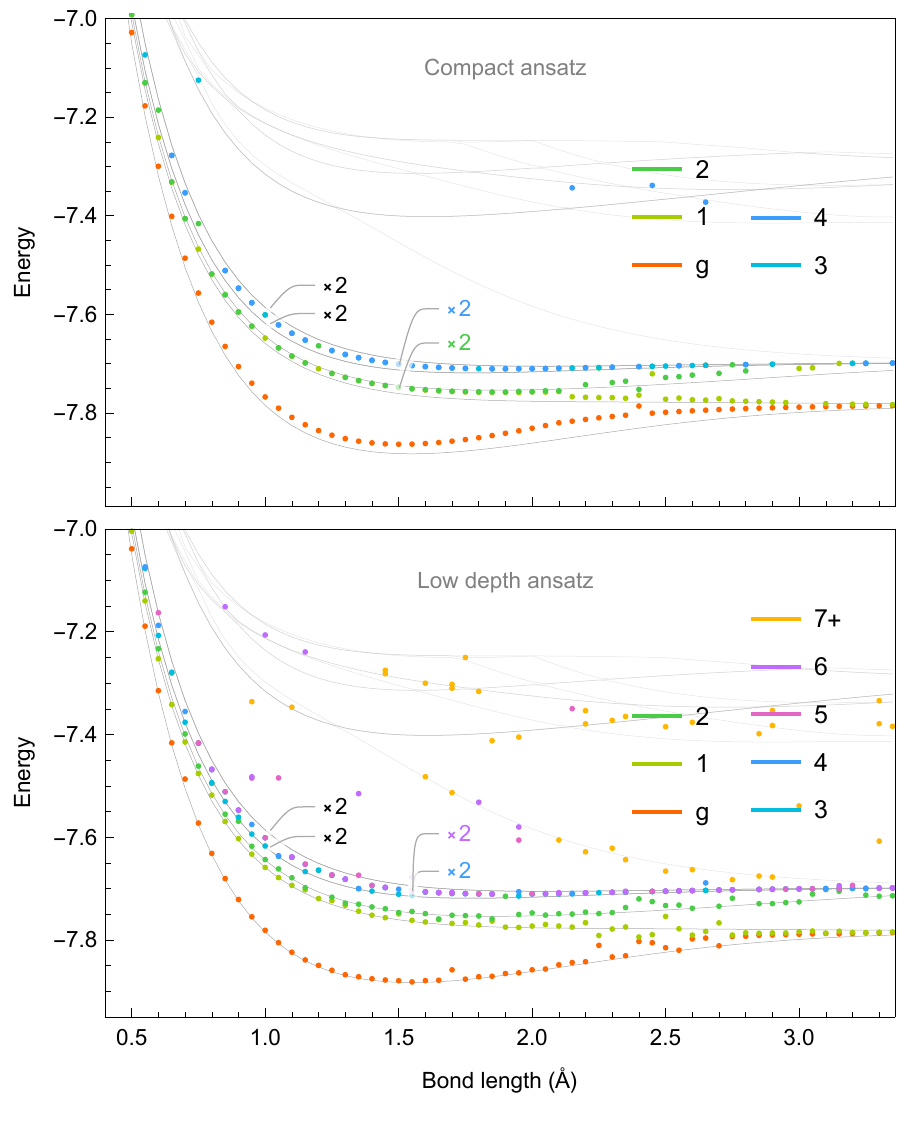}
	\caption{
		The lowest lying states discovered by imaginary time evolution of the 10-qubit LiH Hamiltonian over varying bond length. The gray lines indicate the true spectrum as found by diagonalisation, and the line labels indicate degeneracy in both the true and discovered states. The compact ansatz is used with 70 parameters and for 10k iterations at every bond length. The low depth ansatz uses 145 parameters for 40k iterations. 
	}
	\label{fig:spec_over_bondlength}
\end{figure}

We next task our method with finding several of the lowest lying states of the physical $10$-qubit LiH Hamiltonian as a function of the bond length. The results for both ans\"atze are shown in \figref~\ref{fig:spec_over_bondlength}. The compact ansatz with 70 parameters shows reasonable agreement with the true sepctrum, despite generating only a small submanifold of the full $2^{10}$ Hilbert space. The smooth deviation of the lowest discovered energy with the true ground energy may result from the ansatz's inability to generate the ground state. 
The low depth ansatz with 145 parameters shows a marked improvement in accuracy, and a better discovery of the degenerate energy eigenvalues.

Both ans\"atze show decreased accuracy around bond length $l\approx 2.5$\AA. This was also seen in recent variational eigensolver experiments~\cite{kandala2017hardware}, and attributed to the insufficient power of the low depth ansatz to generate these particularly highly entangled eigenstates~\cite{peruzzo2014variational}.

\section{Discussion}

In this article, we have proposed a variational algorithm for a hybrid quantum-classical computer to discover the spectra of Hamiltonians. 
Our algorithm can also offer a route to enhancing the performace of the ground state solver in Ref.~\cite{mcardle2018variational} since it can eliminate low lying states once found, thus `clearing the way' for a successful identification of the true ground state.
We tested our method on SAT and LiH Hamiltonians, using two different ans\"atze, and successfully obtained estimates of the excitation spectra.
In our simulations we rarely saw variational imaginary time evolution converging to non-eigenstates. In contrast, gradient descent was prone to becoming stuck in local minima which when excited, caused errors in the reported spectrum.

Our results suggest a number of directions for fruitful future research. For instance, how should the necessary ability to accurately generate the energy eigenstates inform the design of the ansatz? 
And, how faithfully must the variational evolution simulate the true imaginary time evolution in order to converge to the lowest lying states?

There are also questions concerning the classical component of the hybrid algorithm. For example, we might seek a fuller understanding of the relationship between the numerical solving algorithm employed by the classical processor and the consequential convergence of variational imaginary time evolution. We elaborate upon this in \appref~\ref{appendix:numerics_details}.

\begin{figure}[t]
	\centering
	\includegraphics[width=\columnwidth]{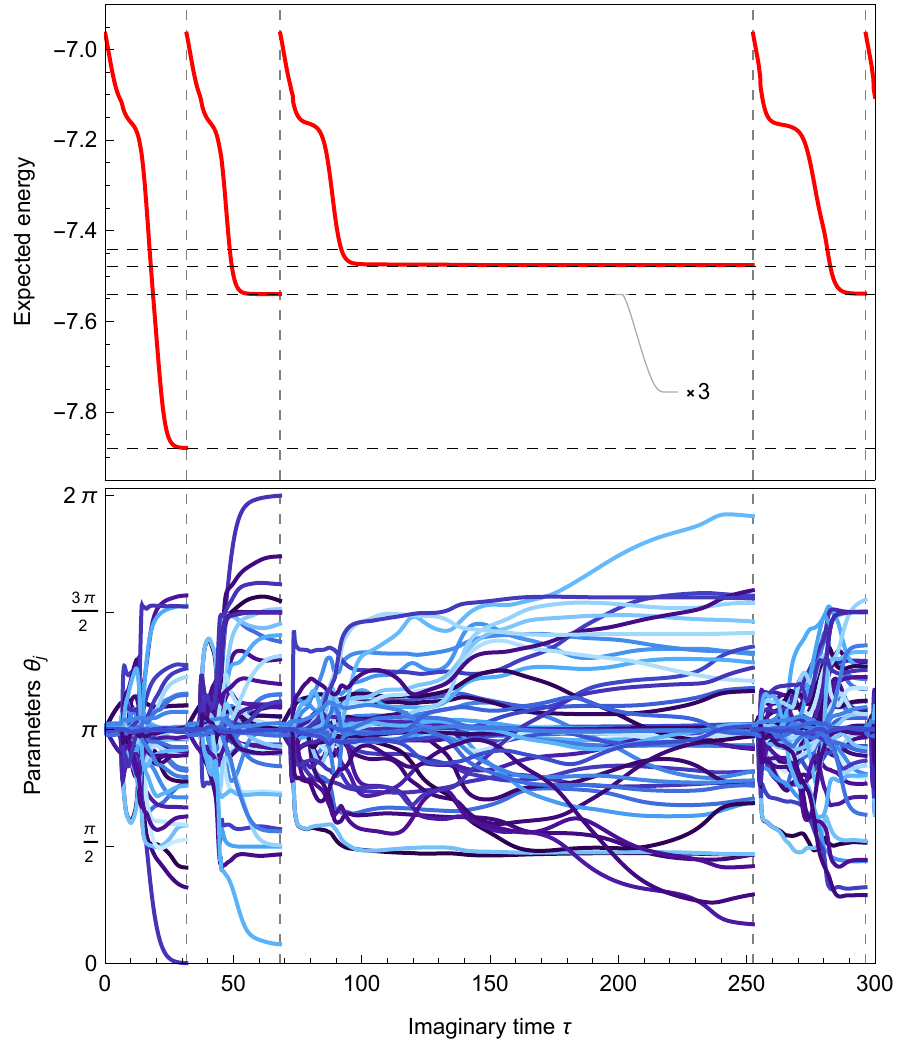}
	\caption{An example of variational imaginary time evolution whereby energy plateaued during a stage of continued parameter change. This is the low depth ansatz of 56 parameters exploring the reduced $6$-qubit LiH Hamiltonian, without parameter re-randomisation.
    }
	\label{fig:false_energy_plateu}
\end{figure}

A final topic to mention is the challenge of {\it predicting} the number of iterations necessary to converge to an eigenstate; \figref~\ref{fig:false_energy_plateu} demonstrates an anomalous simulation where, despite the energy stabilising to an eigenvalue, the parameters continued to change. 

Our method is not limited to exciting eigenstates - we can excite states for which the generating parameters are \textit{a priori} known. This could be applied to eliminate unwanted subspaces from the searched spectrum, such as those which break symmetries or indicate error.
Furthermore, our algorithm can be adapted to modify Hamiltonians in real-time variational simulation~\cite{Li2017}. 
Discovered eigenstates can be excited by different amounts to modulate their new energies, for instance to create or remove energy degeneracies, or create time-dependence in the spectrum. Updating the linear equations in variational simulation by the procedure outlined in this work will then effectively simulate the real-time dynamics under the modified Hamiltonian.
We leave exploring these extensions for a future work.

\section{Acknowledgements}

This work was supported by the EPSRC National Quantum Technology Hub in Networked Quantum Information Technologies. The authors acknowledge the use of the University of Oxford Advanced Research Computing (ARC) facility (\url{http://dx.doi.org/10.5281/zenodo.22558}). 
X. Y. acknowledges support from BP plc. T. J. thanks the Clarendon Fund for their support. S. E. is supported by Japan Student
Services Organization (JASSO) Student Exchange Support
Program (Graduate Scholarship for Degree Seeking
Students).

\bibliographystyle{apsrev4-1}
\bibliography{bibliography}

\appendix

\section{Quantum circuits to obtain the elements of $\mathcal{M}$ and $\mathcal{V}$}
\label{appendix:eval_M_V}

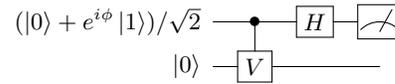
\begin{figure}[htbp!]
\centerline{
\Qcircuit @C=1em @R=.5em {   
\lstick{(\ket{0}+e^{i\phi}\ket{1})/\sqrt{2}}&\ctrl{1}&\gate{H}& \meter\\
\lstick{\ket{0}}&\gate{V}&\qw&\qw
}
}
\caption{A quantum circuit which evaluates ${\Re}(e^{i\phi}\bra{\bar{0}}V\ket{\bar{0}})$. $H$ is the Hadamard gate. The first qubit is measured in the computational $\{\ket{0},\ket{1}\}$ basis, and the average value $\braket{Z}$ (Pauli) of the second qubit equals ${\Re}(e^{i\phi}\bra{\bar{0}}V\ket{\bar{0}})$.}
\label{vmeasure}
\end{figure}

Here we denote the full ansatz unitary as $\mathcal{U}:= U_M(\theta_M) U_{M-1}(\theta_{M-1}) \dots U_1(\theta_1)$, where $U_j$ is the ansatz's $j^\text{th}$ parameterised gate. 
Let $\mathcal{U}_{k,i}$ denote a modification of $\mathcal{U}$ where a new gate $G_{k,i}$ is inserted before the $i^\text{th}$ gate. That is,
\begin{align}
\mathcal{U}_{k,i} := U_M(\theta_M) \dots U_i(\theta_i) G_{k,i} U_{i-1}(\theta_{i-1}) \dots U_1(\theta_1).
\end{align}
We then assume that the derivative $\frac{\partial \ket{\varphi(\vec{\theta}(\tau))}}{\partial \theta_i}$ can be expressed as 
\begin{align}
\frac{\partial \ket{\varphi(\vec{\theta}(\tau))}}{\partial \theta_i}= \sum_k h_{k,i} \, \mathcal{U}_{k,i} \ket{\bar{0}} 
\end{align}
for some family of complex scalars $h_{k,i}$.
We can then express
\begin{align}
\mathcal{M}_{i,j}&= \Re ~\big(\sum_{k,l} h^*_{k,i}h_{l,j} \bra{\bar{0}} \mathcal{U}_{l,i} ^\dag \mathcal{U}_{l,j} \ket{\bar{0}} \big), \\
\mathcal{V}_i &=  \Re \big( \sum_{k, \alpha} h^*_{k,i} f_\alpha \bra{\bar{0}} \mathcal{U}_{k,i} ^\dag P_\alpha \mathcal{U} \ket{\bar{0}}  
 \big),
 \label{eq:qcirc}
\end{align}
where we have expanded the Hamiltonian as a sum of Pauli operators, $H= \sum_\alpha f_\alpha P_\alpha$. Each term in \equref~\eqref{eq:qcirc} can be expressed in the form $c~ \Re (\bra{\bar{0}} e^{i \phi} V \ket{\bar{0}})$ where $V$ is a unitary operator which can be evaluated by using the quantum circuit in \figref~\ref{vmeasure}. 

In reality, far simpler circuits than this controlled $V$ circuit need be implemented. For more detail, refer to Ref~\cite{mcardle2018variational}.

\section{Overlap calculation by using shallow quantum circuit}
\label{appendix:overlap_calc}

\begin{figure}[htbp!]
\includegraphics[width=8cm]{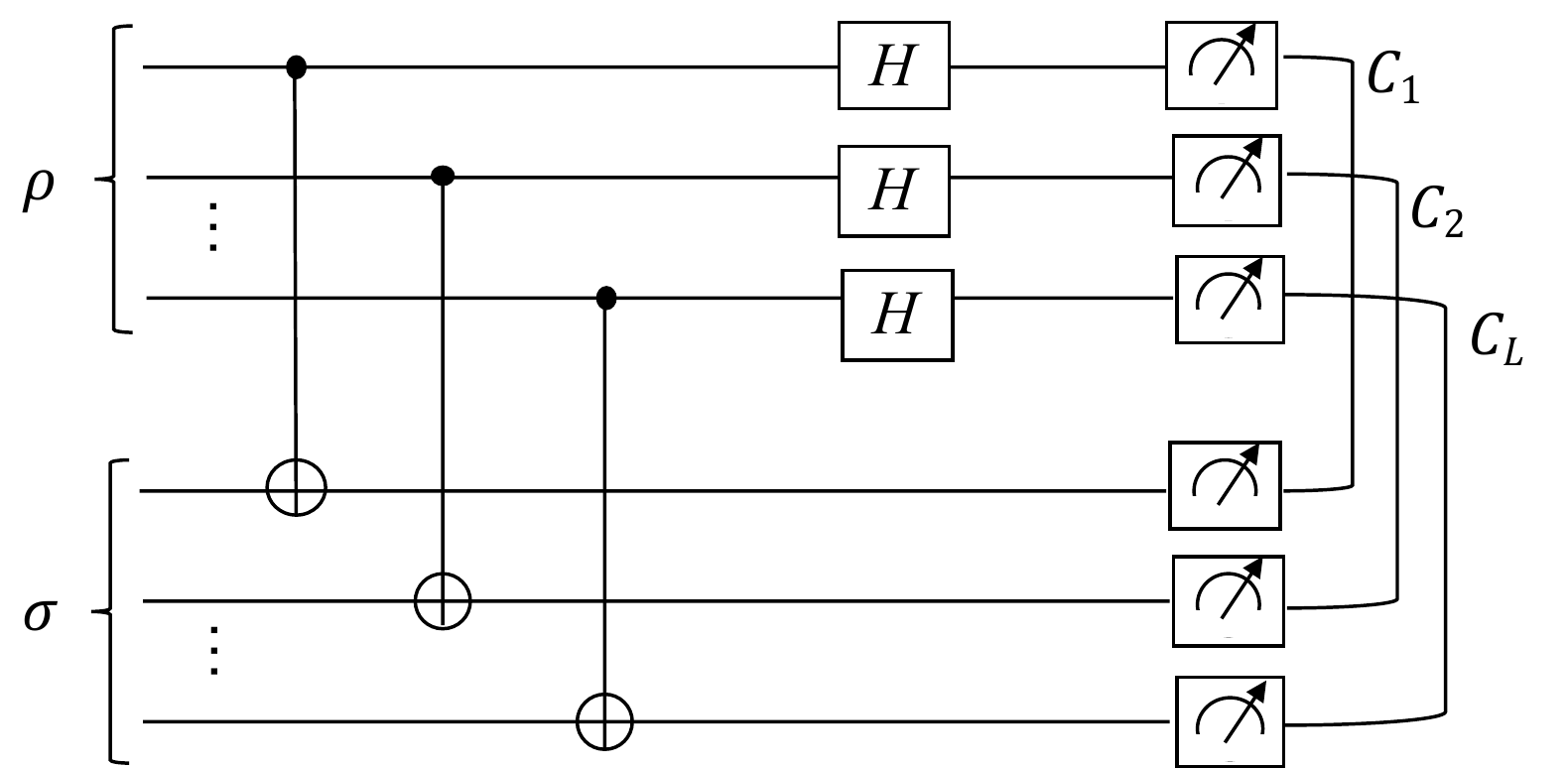}
\caption{Schematic of the shallow swap test circuit from \refref~\cite{shallowswap2}. This circuit evaluates the overlap of two input density matrices $\rho $ and $\sigma$.}
\label{stest}
\end{figure}

Ref.~\cite{shallowswap2} introduces an algorithm for calculating the overlap of two wavefunctions using shallow, constant depth circuits. Interestingly, this algorithm was rediscovered using machine learning \cite{shallowswap1}. We briefly outline the algorithm below, which is visualised in \figref~\ref{stest}.
Let $\rho$ and $\sigma$ denote the two input states, each of $L$ qubits.
We pair each qubit of $\rho$ with one of $\sigma$, applying a controlled-NOT gate between them - controlled on $\rho$ and targeting $\sigma$.
Next, Hadamard gates are applied transversally to the qubits of $\rho$. 
We then measure observable $\bigotimes_{n=1}^L C_n$ where 
\begin{align}
C_n&=\ket{0}\bra{0}_\rho^n \otimes \ket{0}\bra{0}_\sigma^n+\ket{1}\bra{1}_\rho^n \otimes \ket{0}\bra{0}_\sigma^n \\ \nonumber
&+\ket{0}\bra{0}_\rho^n\otimes\ket{1}\bra{1}_\sigma^n-\ket{1}\bra{1}_\rho^n\otimes \ket{1}\bra{1}_\sigma^n.
\end{align}
Here, $\ket{0}\bra{0}^n_\rho$ projects the $n$th qubit of $\rho$ onto the classical $0$ state, and similarly for $\sigma$ and the $1$ state.

Measuring $\bigotimes_{n=1}^L C_n$ is accomplished with post processing, by first measuring each of $C_n$.
If $C_n$ is measured as $\ket{1}\bra{1}_\rho^n\otimes \ket{1}\bra{1}_\sigma^n$, we assign $c_n=-1$, assigning $c_n=1$ for all other outcomes.
The result of observable $\bigotimes_{n=1}^L C_n$ is then $\Pi_{n=1}^L c_n$. By repeating this process and averaging over the results, we can evaluate the overlap function $\mathrm{Tr} (\rho  \sigma)$.

\section{Hamiltonian construction}
\label{appendix:hamil_construction}

\subsection{3SAT}

The boolean satisfiability problem involves finding a satisfying assignment of variables constrained in a propositional formula. For 3SAT, this formula is a set of clauses, each consisting of three terms, which are variables with or without negation. A clause is satisfied by containing at least one true term, and every clause must be simultaneously satisfied to satisfy the formula. Finding a satisfying assignment is NP-complete~\cite{SATdescribe}. We restrict ourselves to 3SAT problems with a single satisfying solution, and map each boolean variable to one qubit - the qubit's $1$ and $0$ classical states correspond to true and false assignments of the variable.
We build a diagonal Hamiltonian from a 3SAT formula by treating each computational basis state as a boolean assignment and energetically penalising it by the number of clauses it fails to satisfy. In this Hamiltonian, the ground-state is the unique solution with zero energy, and the highly-degenerate excited spectrum has integer energies. We test our method on 3SAT Hamiltonians of up to 18 qubits.
We stress that we do not present our method as an efficient 3SAT solver - we instead use 3SATs to construct interesting diagonal Hamiltonians of which our method can discover the spectra.

\subsection{LiH}

We consider the LiH molecule in both a reduced and full spin-orbital basis. We work in the STO-3G basis in which LiH has 12 spin-orbitals: $2 \times ( \{1S\}^H + \{1S, 2P_x, 2P_y, 2P_z\}^{Li} )$.
These 12 orbitals can be mapped to 10 qubits by restriction to non-ionic states with four electrons. For some tests, we additionally reduce LiH to 6 qubits in a reduced-spin orbital basis which has a qualitatively different (and non-physical) spectrum to that in the full basis, though remains interesting for testing our method. 

We reduce the number of active orbitals by first transforming to the natural molecular orbital basis. These are the orbitals which diagonalise the single particle reduced density matrix (1-RDM). We then consider those orbitals with occupation close to unity as being filled, and those orbitals with occupation close to zero as being empty. We can then remove the corresponding fermionic operators from the Hamiltonian. This process is described in greater detail in Refs.~\cite{mcardle2018variational,TrappedIon}.
We then transform our (optionally reduced) fermionic Hamiltonian into a qubit Hamiltonian, using the Bravyi-Kitaev transform~\cite{BravyiKitaev12}. In our six and ten qubit simulations, we have removed two qubits using conservation of electron number and spin~\cite{kandala2017hardware,bravyi2017tapering}. All of these steps were carried out using OpenFermion~\cite{mcclean2017openfermion}, an electronic structure package for quantum computational chemistry.

\section{Implementation of numerical simulations}
\label{appendix:numerics_details}

We simulate the variational imaginary time evolution quantum circuits using the Quantum Exact Simulation Toolkit (QuEST)~\cite{questwhitepaper}.
Direct diagonalisation of the considered Hamiltonians is performed with GSL, which employs a complex form of the symmetric bidiagonalisation and QR reduction method \cite{GslGnuNumericsLibrary,gslDiagAlgorithms}. 

\subsection{Parameter evolution}

We first choose initial parameter values $\vec\theta_0$ which produce a highly excited state in the ansatz circuit. The choice is arbitrary, since random parameters are likely to produce a superposition state with a high expected energy according to the variational principle - our simulations choose $\vec\theta_0$ uniformly randomly in $[0, 2\pi)$. These are fed to an ansatz circuit simulated in QuEST, and the resulting state-vector used to populate $\mathcal{M}$ and $\mathcal{V}$ matrices, which are then fed to GNU Scientific Library (GSL) numerical solving routines~\cite{GslGnuNumericsLibrary}.
We then update the parameters under the variational imaginary time evolution described in Eqs. \eqref{eq:m_and_v_update_equs} and \eqref{eq:update_theta_eq}.

In general, \equref~\eqref{eq:update_theta_eq} can be ill-posed, and direct inversion of $\mathcal{M}$ is numerically unstable.
We instead, after populating $\mathcal{M}$ and $\mathcal{V}$ at every time-step, update the parameters under Tikhonov regularisation, which minimises
\begin{equation}
\| \mathcal{V} - \mathcal{M} \dot{\vec{\theta}} \|^2 + \lambda \| \dot{\vec{\theta}} \|^2
\end{equation}
where the Tikhonov parameter $\lambda$ can be varied to tradeoff accuracy against keeping $\dot{\vec{\theta}}$ small and the parameter evolution smooth. 
Our simulations estimate an ideal $\lambda$ at each time-step by selecting the corner of a 3-point L-curve~ \cite{GslGnuNumericsLibrary,gslDiagAlgorithms}, though force $\lambda \in [10^{-4}, 10^{-2}]$.
This is because too large a $\lambda$ over-restricts the change in the parameters in an iteration and was seen to lead to eventual convergence to non-eigenstates. Meanwhile, no regularisation ($\lambda=0$) saw residuals in $\mathcal{M}^{-1}$ disrupt the monotonic decrease in energy.


Still, using Tikhonov regularisation affords us a larger time-step than other tested methods, which included LU decomposition, least squares minimisation, singular value decomposition (SVD) and truncated SVD. Our simulations typically employ a time-step of $\delta \tau = 10^{-1}$. We suspect the largest stable time-step possible relates to the greatest energy eigenstate with non-negligible probability in the initial ansatz state.

We continue simulating in imaginary time until detecting convergence by a change in the parameters smaller than some threshold for several iterations, typically $\| \Delta \vec{\theta} \| < 10^{-2}$ for $3$.
The parametrised state is then assumed an eigenstate and has its state-vector recorded, to be subsequently excited in the Hamiltonian through modifying $\mathcal{V}$ via \equref~\eqref{eq:v_update_under_excitation} every time-step thereafter. At this point, we reset the parameters to their initial values, restoring the original excited state (or one now of greater energy), and then resume imaginary time evolution. 


\subsection{Populating $\mathcal{M}$ and $\mathcal{V}$}
To save time, our code simulates only the ansatz and Hamiltonian circuits, using several shorcuts to avoid direct simulation of the entire circuits involved in populating $\mathcal{M}$ and $\mathcal{V}$.
Firstly, we calculate each $\partial \ket{\varphi(\vec{\theta}(\tau))}/\partial \theta_i$ term by a fourth-order central finite-difference approximation with a step-size of $\Delta \theta_i = 10^{-5}$, in lieu of simulating the circuits shown in \appref~\ref{appendix:eval_M_V}. Full simulation of these sub-circuits is performed in \refref~\cite{mcardle2018variational}.

$\mathcal{M}$ is then populated by the inner product of these terms, and $\mathcal{V}$ via their inner product with the state-vector produced by simulating the Hamiltonian circuit on the ansatz.
Excitations in $\mathcal{V}$ are introduced merely by the inner product of these terms and the recorded state-vectors of the discovered eigenstates, in lieu of simulating the swap test circuits described in \appref~\ref{appendix:overlap_calc}. Our simulations typically excited the eigenstates by $\alpha \sim 10$, comparable to the gap between the ground and the highest considered excited state of the system.

\end{document}